\documentclass[%
reprint,
superscriptaddress,
amsmath,amssymb,
aps,
prl,
floatfix,
longbibliography
]{revtex4-2}

\usepackage{graphicx}
\usepackage[breaklinks]{hyperref}

\usepackage{times}
\usepackage{dcolumn}
\usepackage{float}
\usepackage[dvipsnames]{xcolor}

\hypersetup{colorlinks=true,linkcolor=blue,citecolor=blue,urlcolor=blue}

\usepackage{mathtools}
\usepackage[normalem]{ulem}

\usepackage[whole]{bxcjkjatype}

\makeatletter
\def\maketitle{
\@author@finish
\title@column\titleblock@produce
\suppressfloats[t]}
\makeatother

\begin{document}


\title{Gyroidal Magnetic Order with Weyl Magnons}

\author{Kiyu Fukui}
\email{k.fukui@aion.t.u-tokyo.ac.jp}
\affiliation{Department of Physical Sciences, Ritsumeikan University, Kusatsu, Shiga 525-8577, Japan}

\author{Yukitoshi Motome}
\affiliation{Department of Applied Physics, The University of Tokyo, Bunkyo, Tokyo 113-8656, Japan}

\date{\today}

\begin{abstract}
The gyroid, a triply periodic minimal surface found widely in nature and materials, has remained largely unexplored as a design principle for magnetic order. We introduce gyroidal magnetic order on the hyperoctagon lattice, the skeletal network of the gyroid, in which each magnetic moment points along the local $C_3$ axis that coincides with the local normal to the gyroid surface. The resulting noncoplanar spin texture constitutes a direct magnetic imprint of the gyroid geometry. We show that this order can be stabilized by single-ion anisotropy, the Dzyaloshinskii-Moriya interaction, or the symmetric off-diagonal $\Gamma$ interaction combined with the Kitaev interaction, and remains robust against quantum fluctuations. Its magnon spectrum hosts symmetry-enforced pseudospin-$1$ and pseudospin-$1/2$ Weyl points that, unlike their electronic counterparts, persist under symmetry-allowed perturbations. Our results establish the gyroid as a geometric platform for magnetic order and symmetry-enforced topological magnon excitations.
\end{abstract}

\maketitle

\begin{figure}
    \centering
     \includegraphics[width=0.9\linewidth,clip]{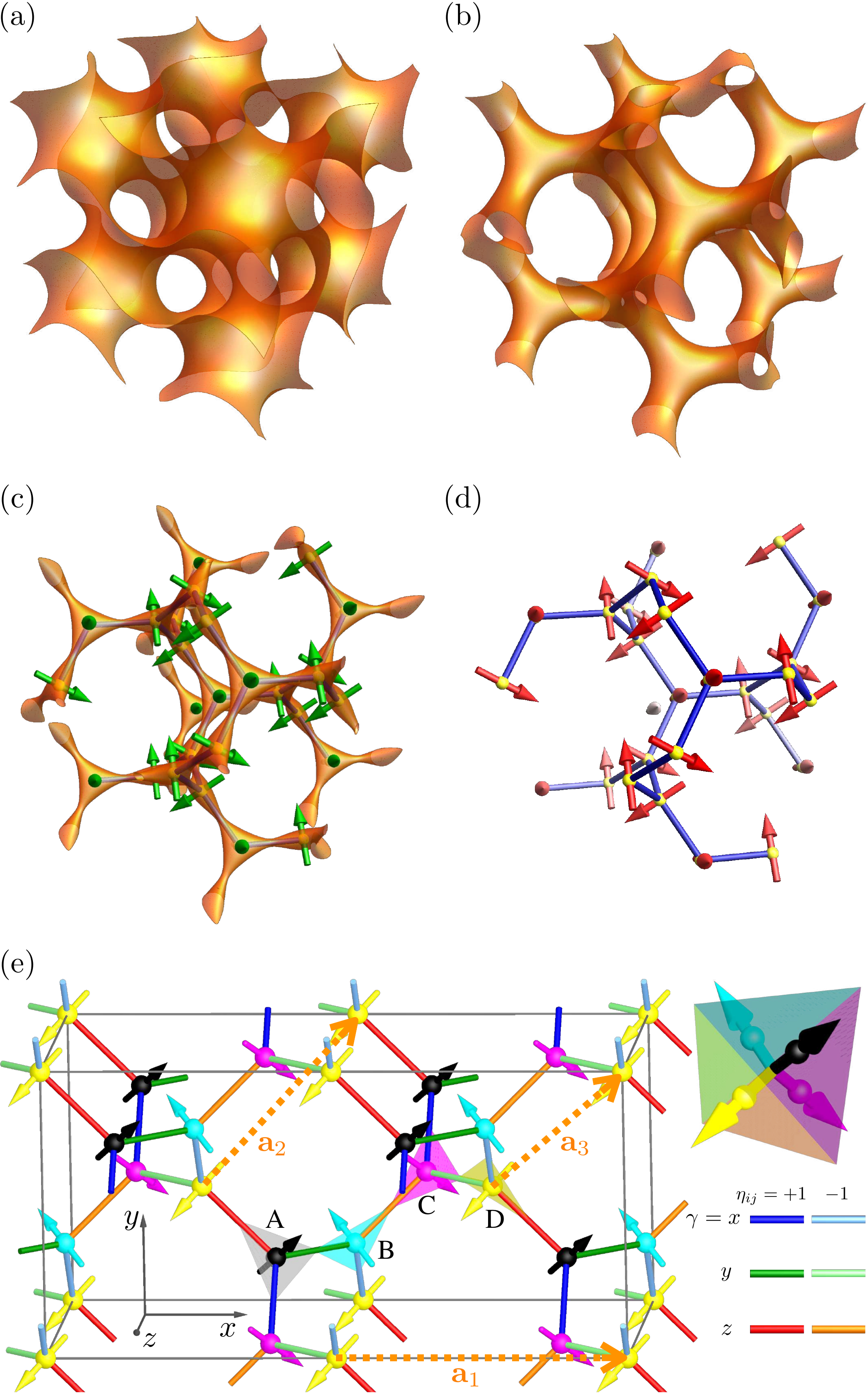}
    \caption{
        Geometry of the gyroid, hyperoctagon lattice, and gyroidal magnetic order. (a) Gyroid surface dividing space into two interpenetrating labyrinthine domains. (b), (c) Surfaces obtained by progressively shrinking one of the two domains toward its skeletal network~\cite{SM}. In (c), the green arrows indicate the surface normals at the points that become lattice sites as the surface shrinks to the limiting discrete structure shown in (d).
        (d) Resulting discrete hyperoctagon lattice, together with a gyroidal magnetic order indicated by red arrows.
        (e) Enlarged view of the gyroidal order in the 16-site conventional unit cell indicated by the black box used in Ref.~\cite{Hermanns2014}. Thick solid arrows indicate the spin directions, which are parallel to the local $C_3$ axes $\mathbf{n}_\eta$ with $\sigma_i=+1$ in Eq.~\eqref{eq:gyroidal}; sites are colored by sublattice A--D. The shaded triangles spanned by the three emanating bonds of each site represent the local bond planes, and correspond to the triangular surface patches remaining during the reduction in (c). The inset illustrates that the four bond planes are parallel to the faces of a regular tetrahedron, with the spins oriented along the corresponding face normals.
        Bond colors distinguish the $x$, $y$, and $z$ bonds and the sign of the $\Gamma$ interaction $\eta_{ij}=\pm 1$ in Eq.~\eqref{eq:JKG_model}. Orange dashed arrows denote the primitive lattice vectors $\mathbf{a}_1$, $\mathbf{a}_2$, and $\mathbf{a}_3$.
    }
    \label{fig:surface2lattice}
\end{figure}

\emph{Introduction.---}
Universality is a central concept in science, referring to the emergence of common phenomena across systems with vastly different microscopic origins and length scales. Its most celebrated manifestation occurs near continuous phase transitions, where microscopically distinct systems exhibit identical critical behavior~\cite{Wilson1974, Fisher1974}. 
A different form of universality is geometric: the recurrence of the same spatial structure across otherwise unrelated systems and length scales. A prominent example is the gyroid [Fig.~\ref{fig:surface2lattice}(a)], a triply periodic minimal surface that divides space into two interpenetrating labyrinthine domains with opposite chiralities. 
First observed in lipid--water systems~\cite{Luzzati1967} and independently introduced as a mathematical minimal surface~\cite{Schoen1970}, the gyroid has since been found in block copolymers~\cite{Hajduk1994,Matsen1994,Schulz1994}, liquid crystals~\cite{Luzzati1968,Hyde1984,Clerc1994}, and biological structures such as butterfly wing nanostructures~\cite{Mille2011}.
Its high stiffness-to-weight ratio and large surface area have also motivated applications in architecture, additive manufacturing, photonic crystals, and biomedical engineering~\cite{Dolan2015,Han2018,Goi2018,Spencer2024,Hossain2025,Korol2026}.
The gyroid thus provides a paradigmatic example of geometric universality.

Remarkably, the continuous gyroid surface possesses a discrete counterpart. Each of the two domains separated by the surface can be continuously reduced to its skeletal graph, the hyperoctagon lattice, a three-dimensional tricoordinated network [Figs.~\ref{fig:surface2lattice}(b)--\ref{fig:surface2lattice}(d)]~\cite{SM}. This continuous reduction establishes a direct correspondence between the continuous gyroid surface and the discrete lattice. Mathematically, the hyperoctagon lattice is the maximal Abelian cover of the complete graph $K_4$ and is distinguished by its strong isotropy~\cite{Sunada2008,Sunada2013}. It has been identified in $K_4$ carbon~\cite{Rignanese2008,Itoh2009}, molecular crystals~\cite{Mizuno2015,Mizuno2019}, metal-organic frameworks (MOFs)~\cite{Mizuno2019,Ishikawa2024,Imajo2025}, and spin-hedgehog lattices~\cite{Kanazawa2017}. Known variously as the Laves graph~\cite{Coxeter1955}, srs net~\cite{OKeeffe2000,Friedrichs2003}, $(10,3)a$ net~\cite{Wells1977}, triamond~\cite{Sequin2008}, diamond twin~\cite{Tagami2014}, $Y^*$ lattice complex~\cite{Koch1978}, and 3/10/c1~\cite{Koch1995}, this network recurs across diverse disciplines, including mathematics, chemistry, and physics.

The hyperoctagon lattice has recently attracted considerable interest in magnetism, particularly stimulated by the synthesis of hyperoctagon cobalt-based MOFs~\cite{Ishikawa2024,Imajo2025} as promising platforms for three-dimensional Kitaev spin liquids~\cite{Hermanns2014}. The Kitaev model is one of the most intensively studied models in quantum magnetism, as it provides an exactly solvable quantum spin liquid~\cite{Kitaev2006}. Its exact solvability, however, relies primarily on the tricoordinated connectivity, and is blind to the underlying gyroid geometry. This raises a fundamental question: Can this geometry be imprinted directly on magnetism? More specifically, can spin-orbit-coupled interactions beyond the ideal Kitaev model stabilize magnetic order that directly inherits this geometry?

In this Letter, we answer these questions affirmatively by introducing \emph{gyroidal magnetic order} on the hyperoctagon lattice. In this order, each magnetic moment points along the local $C_3$ axis coinciding with the normal to the gyroid surface, forming a noncoplanar spin texture that directly inherits the gyroid geometry.
We identify several distinct stabilization mechanisms involving single-ion anisotropy (SIA), the Dzyaloshinskii-Moriya (DM) interaction, and a combination of the Kitaev and symmetric off-diagonal $\Gamma$ interactions, and demonstrate that the resulting order remains robust against quantum fluctuations. Furthermore, its magnetic symmetries enforce pseudospin-1 and pseudospin-1/2 Weyl points in the magnon spectrum, which persist under symmetry-allowed perturbations. 
Our results establish gyroid geometry as a design principle for magnetic order and symmetry-enforced topological magnon excitations.

\emph{Gyroidal order.---}
The hyperoctagon lattice is a tricoordinated chiral lattice with space group
$I4_132$. Its primitive unit cell contains four sites, labeled A--D in Fig.~\ref{fig:surface2lattice}(e). The three neighbors of each site belong to the other three sublattices, so that all four sublattices are mutually connected. 
At each site, the three emanating bonds lie in a common plane, depicted by a shaded triangle in Fig.~\ref{fig:surface2lattice}(e), and are related by a local threefold rotation about the normal to this plane. We refer to the corresponding rotation axis as the local $C_3$ axis, which points along one of the $\langle 111\rangle$ directions. Each local $C_3$ axis intersects the gyroid surface, whose space group is $Ia\bar{3}d$, at an inversion center that interchanges the two labyrinthine domains~\cite{Lord2003}. At each intersection, the threefold symmetry forces the surface normal to be parallel to the local $C_3$ axis. These points are flat points at which the Gaussian curvature vanishes~\cite{Koch1990}. As illustrated in Fig.~\ref{fig:surface2lattice}(c), the flat points become the lattice sites as the surface shrinks, while their surface normals and the remaining triangular surface patches correspond to the spin directions and local bond planes, respectively, shown in Figs.~\ref{fig:surface2lattice}(d) and \ref{fig:surface2lattice}(e).
We choose the corresponding unit vectors as 
$\mathbf n_{\mathrm{A}} =(1, 1, 1)/\sqrt{3}$, $\mathbf n_{\mathrm{B}} =(-1, 1, -1)/\sqrt{3}$, $\mathbf n_{\mathrm{C}} =(1, -1, -1)/\sqrt{3}$, and $\mathbf n_{\mathrm{D}} =(-1, -1, 1)/\sqrt{3}$.
These vectors point toward the vertices of a regular tetrahedron, resulting in $\mathbf n_{\eta}\cdot\mathbf n_{\eta'}=-1/3$ on every nearest-neighbor bond connecting sublattices $\eta$ and $\eta'$, or equivalently, an acute angle of $\cos^{-1}(1/3)\simeq70.5^\circ$ between the corresponding bond planes [Fig.~\ref{fig:surface2lattice}(e)].

We consider spin configurations of the form
\begin{align}\label{eq:gyroidal}
    \mathbf{S}_{i\in\eta}=S\sigma_i\mathbf{n}_{\eta}, \qquad
    \eta\in\{\mathrm{A}, \mathrm{B}, \mathrm{C}, \mathrm{D}\},
\end{align}
where $\mathbf{S}_i$ is the classical spin at site $i$ with $|\mathbf{S}_i|=S$, and $\sigma_i=\pm1$ is a local Ising variable specifying whether the spin is parallel or antiparallel to its local $C_3$ axis. We refer to the two uniform configurations with $\sigma_i=+1$ or $-1$ at every site, which are related by time reversal, as the (uniform) gyroidal order [Fig.~\ref{fig:surface2lattice}(e)].
The gyroidal order belongs to the chiral magnetic space group $I4_1^\prime 32^\prime$, which breaks time-reversal symmetry while its cubic symmetry forbids a net magnetization.
We also define the staggered gyroidal (s-gyroidal) order, in which $\sigma_i\sigma_j=-1$ on every nearest-neighbor bond (this requires a doubled, eight-site magnetic unit cell as detailed in the End Matter). In both orders, the noncoplanar spin textures directly transfer the local geometry of the gyroid surface to the magnetic degrees of freedom.

\begin{figure*}[t]
    \centering
     \includegraphics[width=\linewidth,clip]{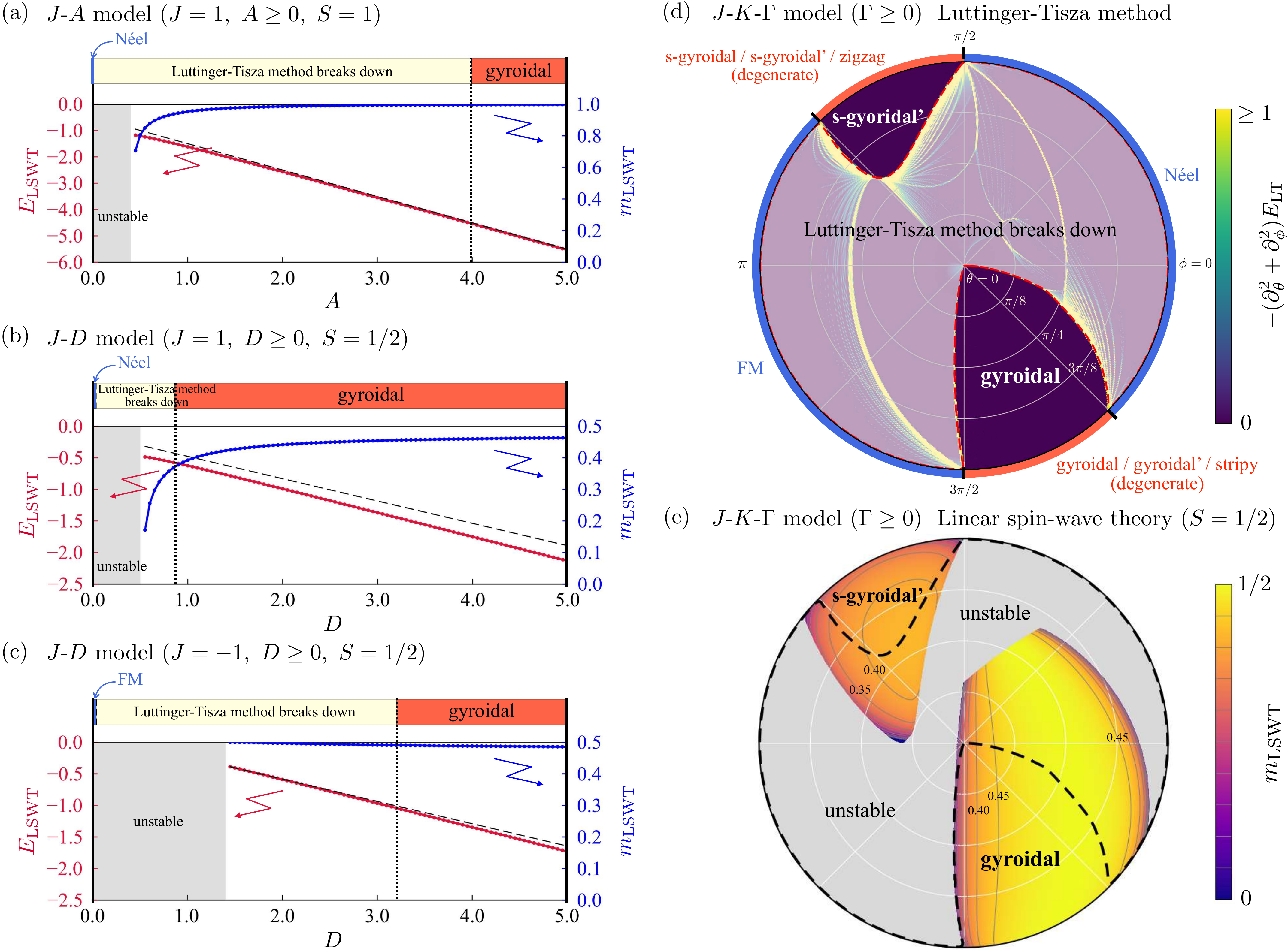}
    \caption{
        Classical phase diagrams and linear spin-wave results. (a) 
        $J$-$A$ model with $J=1$ and $S=1$. (b),(c) $J$-$D$ model with (b) $J=1$ and (c) $J=-1$ for $S=1/2$. The upper strips show the phases obtained from the LT method, while the lower panels display the ground-state energy $E_{\mathrm{LSWT}}$ and ordered moment $m_{\mathrm{LSWT}}$ of the gyroidal order per spin calculated by LSWT. The black vertical dotted lines mark the thresholds above which the LT solution satisfies the local spin-length constraint and yields the gyroidal order. The black dashed curves in the lower panels show the classical energies. 
        In the gray regions, the spin-wave spectrum is unstable.
        (d),(e) Results for the $J$-$K$-$\Gamma$ model with $\Gamma\geq0$ and $S=1/2$, plotted on the upper hemisphere of the parameter sphere: the center corresponds to the strong-$\Gamma$ limit ($\theta=0$), while the rim to $\Gamma=0$ ($\theta=\pi/2$). (d) Second derivative of the LT ground-state energy per spin, $E_{\mathrm{LT}}$, with respect to $\theta$ and $\phi$. (e) Ordered moments of the gyroidal and s-gyroidal' orders per spin obtained from LSWT. The gray regions indicate where neither order is stable within LSWT.
        The regions enclosed by the red and black dashed lines in (d) and (e), respectively, indicate parameter regimes where the LT method breaks down.
    }
    \label{fig:phase_diagram}
\end{figure*}

\emph{Single-ion anisotropy.---}
We first consider the simplest model that realizes the gyroidal order: an antiferromagnetic (AFM) Heisenberg model with an easy-axis SIA along the local $C_3$ axes ($J$-$A$ model),
\begin{equation}
\mathcal H = J\sum_{\langle i,j\rangle}\mathbf S_i\cdot\mathbf S_j
- A\sum_i (\mathbf n_i\cdot\mathbf S_i)^2,
\label{eq:JA_model}
\end{equation}
where $\langle i,j\rangle$ denotes nearest-neighbor pairs, $\mathbf n_i\equiv\mathbf n_\eta$ for $i\in\eta$, and $J>0$ and $A>0$ are the coupling constants of the AFM Heisenberg interaction and the SIA, respectively; the SIA is nontrivial for spin quantum number $S\geq1$.
At $A=0$, the model has a collinear N\'eel ground state on the bipartite hyperoctagon lattice, whereas the SIA favors spin alignment parallel or antiparallel to the local $C_3$ axes. The emergence of the gyroidal order is transparent in the strong-anisotropy limit $A/J\gg1$, where spins are pinned along the directions specified by Eq.~\eqref{eq:gyroidal}. Since $\mathbf{n}_i\cdot\mathbf{n}_j=-1/3$, the exchange energy of each bond becomes $-JS^2\sigma_i\sigma_j/3$. The AFM $J$ thus acts as a ferromagnetic (FM) coupling between the Ising variables $\sigma_i$ and stabilizes the gyroidal order. Conversely, an FM exchange $J<0$ yields an AFM coupling between $\sigma_i$ and stabilizes the s-gyroidal order (see the End Matter).

To assess the stability of the gyroidal order away from the strong-anisotropy limit, we apply the Luttinger-Tisza (LT) method. The result is shown in the upper strip of Fig.~\ref{fig:phase_diagram}(a). At $A=0$, the LT solution yields the N\'eel state. For $0 < A/J \lesssim 4$, it violates the local spin-length constraint, thereby preventing the LT analysis from identifying a physical ground state. For $A/J \gtrsim 4$, the local constraint is again satisfied and the LT solution yields the gyroidal order.

We further examine the effect of quantum fluctuations for $S=1$ using linear spin-wave theory (LSWT), with the resulting ground-state energy $E_{\mathrm{LSWT}}$ and ordered moment $m_{\mathrm{LSWT}}=\lvert\langle \mathbf{S}_i\rangle\rvert\leq S$ shown in the lower panel of Fig.~\ref{fig:phase_diagram}(a).
The ordered moment remains close to its classical value down to $A/J\simeq1$, far below the threshold at which the LT solution recovers the local spin-length constraint, and remains finite until the spin-wave spectrum becomes unstable at $A/J\simeq0.4$. This demonstrates that the gyroidal order is robust against quantum fluctuations over a broad parameter range of $A/J$. The corresponding results for $J<0$, where the s-gyroidal order is
stabilized, are presented in the End Matter.

\emph{Dzyaloshinskii-Moriya interaction.---}
Having established SIA as a direct route to gyroidal magnetism, we next consider the DM interaction, which provides a qualitatively distinct mechanism that remains nontrivial for $S=1/2$. We consider the $J$-$D$ model
\begin{equation}
\mathcal{H} = \sum_{\langle i,j\rangle}\left[J\mathbf S_i\cdot\mathbf S_j
+ D\hat{\mathbf{d}}_{ij}\cdot(\mathbf S_i\times\mathbf S_j)\right],
\label{eq:JD_model}
\end{equation}
where $\hat{\mathbf d}_{ij}=(\mathbf{r}_j-\mathbf{r}_i)/|\mathbf{r}_j-\mathbf{r}_i|$ with $\mathbf{r}_i$ denoting the position of site $i$. Each bond center has two twofold axes perpendicular to the bond, which force the DM vector to be parallel to the bond.
Moreover, since the space group $I4_132$ contains only proper rotations, a single coupling constant $D$ describes all bonds. Here we focus on $D>0$.
Although the DM interaction cannot be optimized on all bonds simultaneously, the gyroidal order achieves the maximum possible energy gain over the entire lattice: specifically, it attains the lower bound $-\sqrt{2}D S^2$ per spin and is therefore an exact classical ground state in the strong-DM limit, unique up to time reversal.

The upper strips in Figs.~\ref{fig:phase_diagram}(b) and \ref{fig:phase_diagram}(c) show the LT results for AFM ($J>0$) and FM ($J<0$) Heisenberg interactions, respectively. In both cases, the LT analysis fails at intermediate $D$, as in the $J$-$A$ model, whereas it yields the gyroidal order at sufficiently large $D$: $D/J\gtrsim0.9$ for $J>0$ and $D/|J|\gtrsim3.2$ for $J<0$. The stabilization of the gyroidal order for both signs of $J$ reflects its unique ground-state character in the strong-DM limit.
The LSWT results shown in the lower panels of Figs.~\ref{fig:phase_diagram}(b) and \ref{fig:phase_diagram}(c) reveal that, even for $S=1/2$, the ordered moment is only weakly reduced from its classical value over a broad range for both signs of $J$, until the spin-wave spectrum becomes unstable at $D/J\simeq0.5$ ($J>0$) and $D/|J|\simeq1.4$ ($J<0$). Reversing the sign of $D$ yields analogous phase diagrams, with the gyroidal order replaced by the s-gyroidal order (see the End Matter).

\emph{Kitaev and $\Gamma$ interactions.---}
Although both SIA and the DM interaction can stabilize gyroidal magnetism, their coupling strengths must be substantial relative to the Heisenberg exchange. We now show that, when combined with the Kitaev interaction, an infinitesimal symmetric off-diagonal $\Gamma$ interaction suffices to stabilize the gyroidal order. We consider the Heisenberg-Kitaev-$\Gamma$ ($J$-$K$-$\Gamma$) model, which also serves as a minimal model for the hyperoctagon MOF Kitaev materials~\cite{Ishikawa2024,Imajo2025}:
\begin{align}
\mathcal H \!=\! \sum_{\gamma}\!\!\sum_{\langle i,j\rangle_\gamma}\!\!
\Big[
J\mathbf S_i\cdot\mathbf S_j
+K S_i^\gamma S_j^\gamma +\eta_{ij}\Gamma\big(S_i^\alpha S_j^\beta+S_i^\beta S_j^\alpha\big)
\Big],
\label{eq:JKG_model}
\end{align}
where $\langle i,j\rangle_\gamma$ denotes nearest-neighbor pairs on $\gamma$ bonds, $(\alpha,\beta,\gamma)$ is a permutation of $(x,y,z)$, and $\eta_{ij}=\pm1$ is a bond-dependent sign fixed by the site $C_3$ and bond $C_2$ symmetries [Fig.~\ref{fig:surface2lattice}(e)]; this sign structure of the $\Gamma$ interaction encodes the underlying gyroid geometry. 
We parametrize the couplings as $J=\sin\theta\cos\phi$, $K=\sin\theta\sin\phi$, and $\Gamma=\cos\theta$. For classical spins, flipping all spins in one of the two partite sets maps $(J,K,\Gamma)$ onto $(-J,-K,-\Gamma)$ and the gyroidal order onto the s-gyroidal order; we thus present the results for $\Gamma\geq0$ and defer those for $\Gamma<0$ to the End Matter.

Figure~\ref{fig:phase_diagram}(d) shows the LT results. Unlike on the two-dimensional honeycomb~\cite{Rau2014,Kruger2020,Fukui2026} and three-dimensional hyperhoneycomb~\cite{Lee2016,Kruger2020} lattices, the LT analysis fails over large portions of the parameter space. Where applicable, it yields the gyroidal and s-gyroidal' orders; the latter is obtained from the s-gyroidal order by exchanging the spin directions between two pairs of sublattices and rotating the moments [see Fig.~\ref{fig:spinlattice}(d) in the End Matter]. 
Steepest-descent energy minimization over the full parameter space confirms that these two gyroidal orders remain stable over broad regions.

The origin of the gyroidal phase is clearest near $\Gamma=0$ ($\theta=\pi/2$), corresponding to the rim in Fig.~\ref{fig:phase_diagram}(d). At $\Gamma=0$, the gyroidal, gyroidal', and stripy states are classically degenerate for $3/2\leq\phi/\pi\leq7/4$, while the s-gyroidal, s-gyroidal', and zigzag states are degenerate for $1/2\leq\phi/\pi\leq3/4$. Here, the gyroidal' state is obtained from the gyroidal order in the same way as the s-gyroidal' order [see Fig.~\ref{fig:spinlattice}(c) in the End Matter]. The classical energies per spin of the first three states are
\begin{align}
E_{\mathrm{gyroidal}}/S^2 &=-\tfrac{1}{2}(J-K)-\Gamma, \label{eq:E_gyroidal}\\
E_{\mathrm{gyroidal}'}/S^2 &=-\tfrac{1}{2}(J-K)+\tfrac{1}{2}\Gamma, \label{eq:E_gyroidalprime}\\
E_{\mathrm{stripy}}/S^2 &=-\tfrac{1}{2}(J-K). \label{eq:E_stripy}
\end{align}
An arbitrarily small positive $\Gamma$ thus lifts the degeneracy in favor of the gyroidal order, which maximizes the energy gain from the $\Gamma$ interaction: in the strong-$\Gamma$ limit ($\theta=0$), corresponding to the center of Fig.~\ref{fig:phase_diagram}(d), it attains the lower bound $-\Gamma S^2$ per spin. This explains why the gyroidal phase extends from the rim at $\Gamma=0$ to the strong-$\Gamma$ limit at the center.

As in the preceding two models, we assess quantum fluctuations for $S=1/2$ using LSWT. At $\Gamma=0$, quantum order by disorder selects the stripy and zigzag states from the degenerate manifolds.
For any $\Gamma>0$, however, these collinear states are no longer classical stationary states, whereas the gyroidal and s-gyroidal' orders acquire an energy gain of $O(\Gamma S^2)$; within the $1/S$ expansion, this classical energy gain dominates the $O(\sqrt{J^2+K^2+\Gamma^2}S)$ quantum correction. As shown in Fig.~\ref{fig:phase_diagram}(e), the ordered moments of the gyroidal and s-gyroidal' orders are only weakly reduced over broad regions and remain finite well beyond the boundaries of the LT-accessible regions. These results demonstrate that both orders are robust against quantum fluctuations.

\begin{figure}[t]
    \centering
     \includegraphics[width=\linewidth,clip]{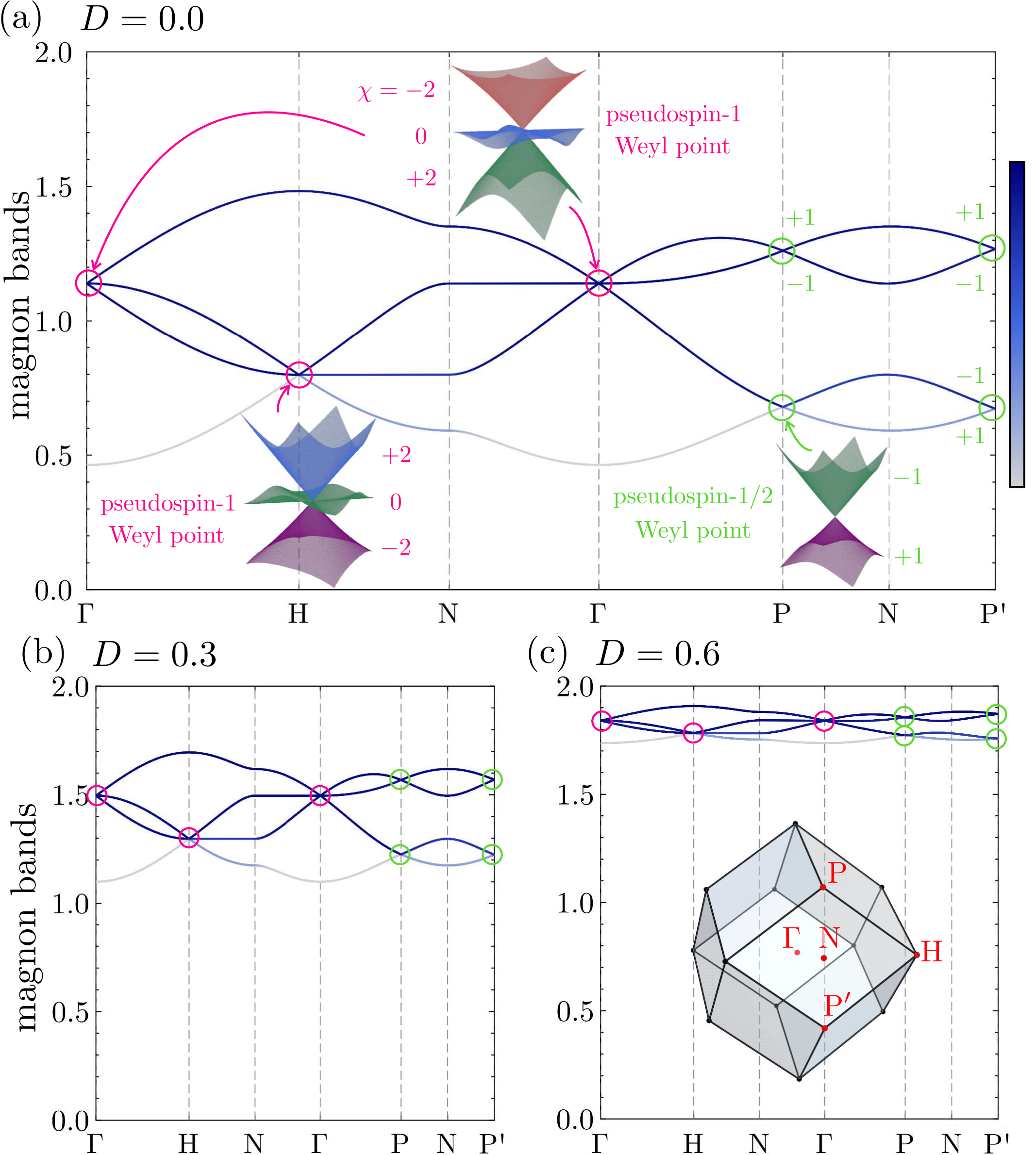}
    \caption{
        Magnon band structures of the gyroidal order in the $J$-$K$-$\Gamma$ model at $(\theta/\pi,\phi/\pi)=(0.40,1.70)$, with the DM interaction of (a) $D=0$, (b) $D=0.3$, and (c) $D=0.6$. Band colors indicate the spectral weight of each magnon mode in the zero-temperature dynamical
        spin structure factor (summed over spin components) obtained by LSWT. Pink and green circles mark the pseudospin-$1$ and pseudospin-$1/2$ Weyl points, respectively, and the integers denote the monopole charges $\chi$ of the corresponding bands. The insets in (a) show the band dispersions in the $(\delta k_x,\delta k_z)$ plane near the Weyl points. The inset in (c) shows the first Brillouin zone with the high-symmetry points. 
    }
    \label{fig:magnon_bands}
\end{figure}

\emph{Magnon bands with Weyl points.---}
Finally, we examine the magnon spectrum of the gyroidal order obtained by LSWT. Figure~\ref{fig:magnon_bands}(a) shows the results at $(\theta/\pi,\phi/\pi)=(0.40,1.70)$, a representative point in the gyroidal phase of the $J$-$K$-$\Gamma$ model in Eq.~\eqref{eq:JKG_model}. The spectrum closely resembles the tight-binding band structure of electrons on the hyperoctagon lattice~\cite{Manes2012,Tsuchiizu2016,Yoshida2026}: three bands cross linearly at the $\Gamma$ and $\mathrm{H}$ points, while two bands cross linearly at the $\mathrm{P}$ and $\mathrm{P^\prime}$ points. Integrating the Berry curvature of each band over a closed surface enclosing a crossing yields monopole charges~\cite{Vanderbilt2018}, $\chi=-2$, $0$, and $+2$ for the three bands at the $\Gamma$ and $\mathrm{H}$ points, and $\chi=\pm1$ at the $\mathrm{P}$ and $\mathrm{P^\prime}$ points [Fig.~\ref{fig:magnon_bands}(a)]. These crossings therefore constitute pseudospin-$1$ Weyl points at $\Gamma$ and $\mathrm{H}$ and pseudospin-$1/2$ Weyl points at $\mathrm{P}$ and $\mathrm{P^\prime}$~\cite{Bradlyn2016}.

These Weyl points are enforced by the $I4_1^\prime32^\prime$ magnetic symmetry, whose unitary subgroup $I2_13$ protects the degeneracies at these high-symmetry points~\cite{Bradley2009,Cano2019,Corticelli2023}. In contrast to the electronic case~\cite{Manes2012,Tsuchiizu2016,Yoshida2026}, the $\mathrm{P}$ and $\mathrm{P^\prime}$ points are not related by symmetry because time-reversal symmetry is broken. The corresponding magnon Weyl points therefore occur at slightly different energies, resulting in a nonreciprocal magnon dispersion. Since these properties are dictated solely by the magnetic space group, they persist beyond LSWT upon including magnon-magnon interactions, provided that the gyroidal order and its symmetry are preserved, although magnon damping may broaden the spectral features.

In the electronic counterpart, the spin degree of freedom doubles the bands, and spin-orbit coupling splits the resulting sixfold degeneracy at the $\Gamma$ point into pseudospin-$3/2$ and pseudospin-$1/2$ Weyl points~\cite{Tsuchiizu2016}. The magnon bands of the gyroidal order have no such spin degeneracy. Moreover, the local $C_3$ symmetry forces the local field at each site to be parallel to the corresponding $C_3$ axis. Consequently, any symmetry-allowed interaction neither cants the moments nor lowers the magnetic symmetry. The Weyl points therefore persist as long as the gyroidal order remains stable. We demonstrate this robustness in Figs.~\ref{fig:magnon_bands}(b) and \ref{fig:magnon_bands}(c) by adding the DM interaction in Eq.~\eqref{eq:JD_model}, the only symmetry-allowed nearest-neighbor DM term and a natural spin-orbit-induced perturbation, to the $J$-$K$-$\Gamma$ model. Upon further increasing $D$, band inversions occur at $\Gamma$ and H at $D=D_{\mathrm c}({\Gamma})\simeq0.705$ and $D_{\mathrm c}({\mathrm H})\simeq0.738$, respectively. These band inversions, together with additional nearby band touchings, exchange the monopole-charge configurations at the two points. Details of these topological transitions will be reported elsewhere.

\emph{Conclusions.---}
We have introduced gyroidal magnetic order on the hyperoctagon lattice, in which each magnetic moment points along the local $C_3$ axis normal to the gyroid surface, generating a noncoplanar spin texture that directly imprints the gyroid geometry. This order can be stabilized through distinct spin-orbit-coupled mechanisms, including SIA, DM, and combined Kitaev and $\Gamma$ interactions. In each case, the gyroidal order maximizes the energy gain from the corresponding interactions and remains robust against quantum fluctuations over broad parameter regions. The magnon spectrum hosts symmetry-enforced pseudospin-$1$ and pseudospin-$1/2$ Weyl points, together with a nonreciprocal dispersion. Since the local $C_3$ symmetry prevents any symmetry-allowed interaction from canting the moments, these Weyl points persist, in contrast to their electronic counterparts. Our results establish gyroid geometry as a platform for magnetic order and symmetry-enforced topological magnons.

The broad range of stabilization mechanisms suggests that gyroidal order may be realized in a variety of spin-orbit-coupled hyperoctagon magnets. Recently synthesized hyperoctagon MOF magnets~\cite{Ishikawa2024,Imajo2025} provide particularly promising candidates. Once gyroidal order is experimentally identified, its magnon spectrum and associated Weyl points could be directly probed by inelastic neutron scattering. The nonreciprocal and topological character of the magnons may also give rise to distinctive transport phenomena, such as nonreciprocal magnon transport and a field-induced magnon thermal Hall effect. As the gyroidal phase lies adjacent to the FM Kitaev limit of the $J$-$K$-$\Gamma$ model, its competition with the Kitaev quantum spin liquid, as well as quantum effects beyond LSWT, deserves further investigation.

\begin{acknowledgments}
We thank H.~Ishikawa, H.~Iwahata, R.~Okuma, H.~Takeda, and T.~Wada for fruitful discussions. K.~F. also thanks M.~Tsuchiizu for useful discussions and T.~Yoneda for making tangible models of the gyroid surface using a 3D printer.
Parts of the numerical calculations have been done using the facilities of the Supercomputer Center, the Institute for Solid State Physics, the University of Tokyo, the Information Technology Center, the University of Tokyo, and the Center for Computational Science, University of Tsukuba.
This work was supported by the Japan Society for the Promotion of Science (JSPS) KAKENHI Grant Nos. JP20H00122, JP24K17009, JP25H01247, and JP26H02011.
\end{acknowledgments}

\bibliography{library}

\clearpage
\appendix
\onecolumngrid
\begin{center}
{\bfseries END MATTER}
\end{center}
\twocolumngrid

\begin{figure*}[h]
    \centering
     \includegraphics[width=0.85\linewidth,clip]{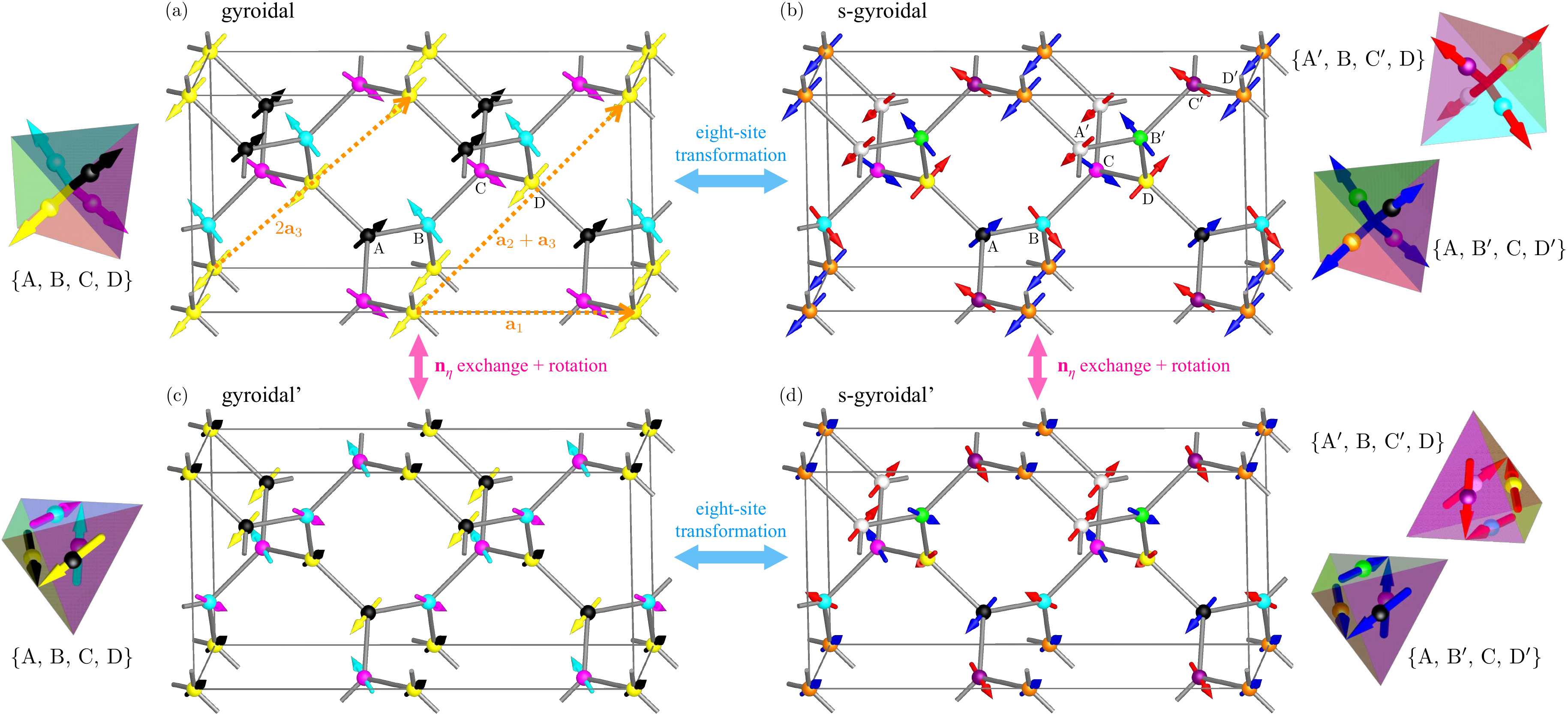}
    \caption{
        Spin configurations of the (a) gyroidal, (b) s-gyroidal, (c) gyroidal', and (d) s-gyroidal' orders and the relations among them. In (a) and (c), sites are colored according to the four sublattices A--D, while in (b) and (d), they are colored according to the eight sublattices A--D and A$'$--D$'$ of the eight-site unit cell extended along $\mathbf a_3$. The configurations in (c) and (d) are given by Eq.~\eqref{eq:gyroidal_prime} with $P:\mathrm A\leftrightarrow\mathrm D,\ \mathrm B\leftrightarrow\mathrm C$ and $\beta=0$. In (a) and (c), $\sigma_i=+1$ at every site, whereas in (b) and (d), blue and red arrows denote spins with $\sigma_i=+1$ and $-1$, respectively. 
        In (c), arrows are colored according to the sublattice $P(\eta)$, indicating the exchange of $\mathbf n_\eta$. 
        As in Fig.~\ref{fig:surface2lattice}(e), the insets show the spin directions relative to the faces of a regular tetrahedron parallel to the corresponding bond planes, separately for the two partite sets in (b) and (d); the spins are normal to the faces in (a) and (b) and lie within them in (c) and (d).
        Other notations are common to those in Fig.~\ref{fig:surface2lattice}(e).
    }
    \label{fig:spinlattice}
\end{figure*}

\emph{Eight-site transformation and staggered gyroidal order.---}
Here we introduce the eight-site transformation, which relates the gyroidal and s-gyroidal orders, as well as the gyroidal' and s-gyroidal' orders, as shown in Fig.~\ref{fig:spinlattice}. Although the primitive unit cell of the hyperoctagon lattice contains four sublattices A--D as shown in Figs.~\ref{fig:surface2lattice}(e) and \ref{fig:spinlattice}(a), each of them is connected to the other three. Consequently, the bipartite structure of the lattice cannot be defined within this unit cell. We therefore consider an enlarged unit cell obtained by extending the primitive unit cell by one period along $\mathbf{a}_3$, which contains eight sublattices A--D and A$'$--D$'$ as shown in Fig.~\ref{fig:spinlattice}(b), with primitive translation vectors $\mathbf{a}_1$, $\mathbf{a}_2+\mathbf{a}_3$, and $2\mathbf{a}_3$. In this enlarged unit cell, the lattice is divided into two partite sets, \{A, B$'$, C, D$'$\} and \{A$'$, B, C$'$, D\}, with every bond connecting sites in different sets. 

The eight-site transformation reverses all spins in one of the two partite sets, say \{A$'$, B, C$'$, D\}, while leaving those in the other set unchanged. Applied to the gyroidal order defined in Eq.~\eqref{eq:gyroidal} [Fig.~\ref{fig:spinlattice}(a)], in which $\sigma_i=+1$
at every site, it reverses $\sigma_i$ only in that set. Since every bond connects the two sets, the resulting configuration satisfies $\sigma_i\sigma_j=-1$ on every bond, and hence corresponds to the s-gyroidal order [Fig.~\ref{fig:spinlattice}(b)].
Likewise, the transformation maps the gyroidal' order onto the s-gyroidal' order, as detailed below [Figs.~\ref{fig:spinlattice}(c) and \ref{fig:spinlattice}(d)], the FM state onto the N\'eel state, and the stripy state onto the zigzag state appearing in Fig.~\ref{fig:phase_diagram}.

Since every bond connects sites in different partite sets, this transformation reverses the sign of every nearest-neighbor bilinear interaction, while leaving the SIA invariant. For classical spins, it thus maps the $J$-$A$ model with $(J, A)$ onto that with $(-J, A)$, the $J$-$D$ model with $(J, D)$ onto that with $(-J, -D)$, and the $J$-$K$-$\Gamma$ model with $(J, K, \Gamma)$ onto that with $(-J, -K, -\Gamma)$. The classical phase diagrams for $J<0$ in the $J$-$A$ model, $D<0$ in the $J$-$D$ model, and $\Gamma<0$ in the $J$-$K$-$\Gamma$ model therefore follow directly from Fig.~\ref{fig:phase_diagram}, with the orders replaced by their counterparts. 

\begin{figure*}
    \centering
     \includegraphics[width=0.85\linewidth,clip]{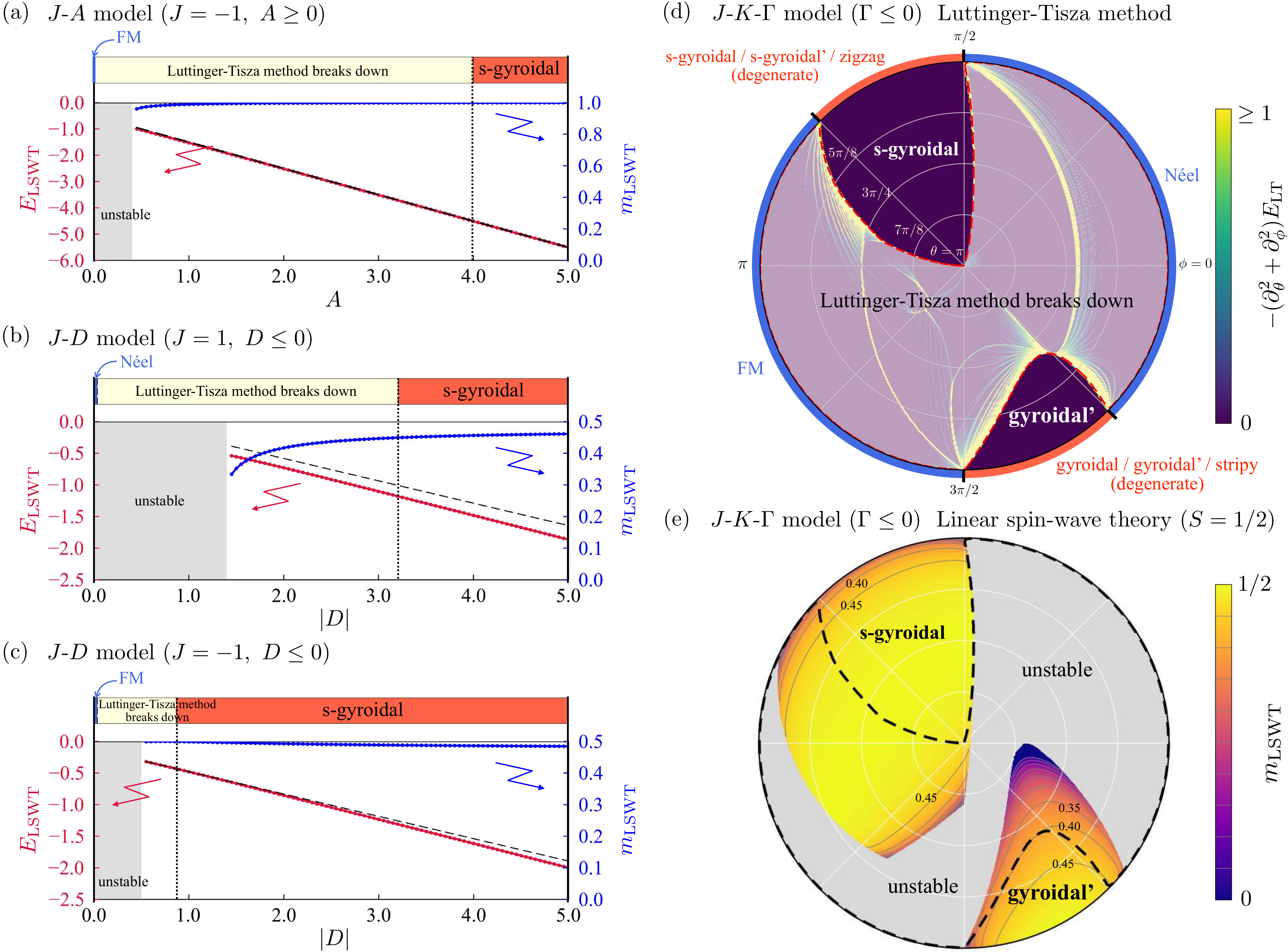}
    \caption{
        Classical phase diagrams and linear spin-wave results for the parameter regions related to those in Fig.~\ref{fig:phase_diagram} by the eight-site transformation. 
        (a) $J$-$A$ model with $J=-1$ and $S=1$. (b),(c) $J$-$D$ model with $D\leq 0$ for (b) $J=1$ and (c) $J=-1$, with $S=1/2$.
        (d),(e) Results for the $J$-$K$-$\Gamma$ model with $\Gamma\leq0$ and $S=1/2$, plotted on the lower hemisphere of the parameter sphere centered at $\theta=\pi$.
        The notations are common to those in Fig.~\ref{fig:phase_diagram}, with the gyroidal and s-gyroidal' orders replaced by the s-gyroidal and gyroidal' orders, respectively.
    }
    \label{fig:phase_diagram_staggered}
\end{figure*}

Figure~\ref{fig:phase_diagram_staggered} shows the results for the parameter regions related to those in Fig.~\ref{fig:phase_diagram} by the eight-site transformation. As expected, the LT results are identical to those in Fig.~\ref{fig:phase_diagram}, apart from the replacement of the orders by their counterparts: the s-gyroidal order appears for $J<0$ in the $J$-$A$ model [Fig.~\ref{fig:phase_diagram_staggered}(a)] and for $D<0$ in the $J$-$D$ model [Figs.~\ref{fig:phase_diagram_staggered}(b) and \ref{fig:phase_diagram_staggered}(c), corresponding to Figs.~\ref{fig:phase_diagram}(c) and \ref{fig:phase_diagram}(b), respectively], and the s-gyroidal and gyroidal' orders appear for $\Gamma<0$ in the $J$-$K$-$\Gamma$ model [Fig.~\ref{fig:phase_diagram_staggered}(d), corresponding to Fig.~\ref{fig:phase_diagram}(d)]. The thresholds for spin-wave instability also coincide with those in Fig.~\ref{fig:phase_diagram}, since they are determined by the classical stability of the ordered states, which is invariant under the transformation. 
In contrast, the ordered moments differ, since the transformation does not preserve the spin commutation relations for quantum spins. Nevertheless, the ordered moments of the s-gyroidal order in Figs.~\ref{fig:phase_diagram_staggered}(a)--\ref{fig:phase_diagram_staggered}(c) and \ref{fig:phase_diagram_staggered}(e), as well as that of the gyroidal' order in Fig.~\ref{fig:phase_diagram_staggered}(e), are only weakly reduced and remain finite over wide parameter regions, demonstrating that these counterparts of the gyroidal and
s-gyroidal' orders are also robust against quantum fluctuations.

\emph{Gyroidal' and s-gyroidal' orders.---}
The gyroidal' and s-gyroidal' orders are obtained from the gyroidal and s-gyroidal orders, respectively, by exchanging $\mathbf n_\eta$ in Eq.~\eqref{eq:gyroidal} between two pairs of sublattices and rotating the spins so that each lies parallel to its local bond plane. 
Let $P$ denote such an exchange, e.g., $P:\mathrm A\leftrightarrow\mathrm D,\ \mathrm B\leftrightarrow\mathrm C$. The construction proceeds in three steps.
(i) Replacing $\mathbf n_\eta$ by $\mathbf n_{P(\eta)}$ in Eq.~\eqref{eq:gyroidal} gives $\mathbf S_{i\in\eta}=S\sigma_i\mathbf n_{P(\eta)}$, such that each spin points along the local $C_3$ axis of its partner sublattice and makes an angle of $\cos^{-1}(-1/3)\simeq109.5^\circ$ with its own axis $\mathbf n_\eta$.
(ii) Each spin is rotated toward $\mathbf n_\eta$, within the plane spanned by $\mathbf n_{P(\eta)}$ and $\mathbf n_\eta$, by $\sin^{-1}(1/3)\simeq19.5^\circ$, so that it becomes perpendicular to $\mathbf n_\eta$, i.e., parallel to its local bond plane.
The resulting direction is the normalized projection of $\mathbf n_{P(\eta)}$ onto this
plane,
\begin{align}\label{eq:e_eta}
    \mathbf e_\eta
    =\frac{\mathbf n_{P(\eta)}-(\mathbf n_{P(\eta)}\cdot\mathbf n_\eta)\,\mathbf n_\eta}
          {\lvert\mathbf n_{P(\eta)}-(\mathbf n_{P(\eta)}\cdot\mathbf n_\eta)\,\mathbf n_\eta\rvert}
    =\frac{3\mathbf n_{P(\eta)}+\mathbf n_\eta}{2\sqrt2}.
\end{align}
(iii) Since any further rotation of the spins about their local $C_3$ axes keeps them within
the bond planes, the gyroidal' and s-gyroidal' orders can be written more generally as
\begin{align}\label{eq:gyroidal_prime}
    \mathbf S_{i\in\eta}=S\sigma_i\left[\cos\beta\,\mathbf e_\eta
    +\sin\beta\,\mathbf n_\eta\times\mathbf e_\eta\right],
\end{align}
where $\beta$ is a rotation angle common to all sites, and $\sigma_i$ is uniform for the gyroidal' order and staggered for the s-gyroidal' order. The configurations shown in Figs.~\ref{fig:spinlattice}(c) and \ref{fig:spinlattice}(d) correspond to $\beta=0$.
The three possible exchanges, $(\mathrm{A}\leftrightarrow\mathrm{D},\mathrm{B}\leftrightarrow\mathrm{C})$, $(\mathrm{A}\leftrightarrow\mathrm{C},\mathrm{B}\leftrightarrow\mathrm{D})$, and $(\mathrm{A}\leftrightarrow\mathrm{B},\mathrm{C}\leftrightarrow\mathrm{D})$ originate from the threefold symmetry and correspond to $\beta=0$, $2\pi/3$, and $-2\pi/3$, respectively, within the continuous family in Eq.~\eqref{eq:gyroidal_prime}, while $\beta\to\beta+\pi$ corresponds to time reversal. 

An analytical calculation shows that the classical energy of Eq.~\eqref{eq:JKG_model} for the gyroidal' order is independent of $\beta$ and given by Eq.~\eqref{eq:E_gyroidalprime}. This energy is lower than those of the gyroidal and stripy states for $\Gamma<0$; correspondingly, the s-gyroidal' order is favored for $\Gamma>0$, as follows from the eight-site transformation. The independence of the classical energy from $\beta$ gives rise to an accidental continuous degeneracy. For any $\beta$, the configurations at $\beta$ and $\beta\pm2\pi/3$ are related by threefold symmetry, while the configuration at $\beta+\pi$ is related by time reversal. Quantum fluctuations lift the accidental continuous degeneracy through order by disorder, with LSWT selecting $\beta=0$, $\pm2\pi/3$, and their time-reversal counterparts over wide parameter regions. We therefore use $\beta=0$ as a representative configuration for the s-gyroidal' and gyroidal' orders in Figs.~\ref{fig:phase_diagram}(e) and \ref{fig:phase_diagram_staggered}(e), respectively, as well as in Figs.~\ref{fig:spinlattice}(c) and \ref{fig:spinlattice}(d).

\end{document}